\documentclass[aps,pra,twocolumn,superscriptaddress]{revtex4-1}

\usepackage{amsmath}
\usepackage{bm}
\usepackage{hyperref}
\usepackage{times}
\usepackage{amsmath,bm}
\usepackage[dvips]{graphicx}
\usepackage{amsmath,amssymb,amsthm,mathrsfs,amsfonts,dsfont}
\usepackage{subfigure, epsfig}
\usepackage{braket}
\usepackage{ulem}
\usepackage{bm}
\usepackage{enumerate}
\usepackage{color}
\hypersetup{colorlinks=true,citecolor=blue,linkcolor=blue,urlcolor=blue}

\begin{document}

\title{Vector DC magnetic-field sensing 
with reference microwave field 
using perfectly aligned nitrogen-vacancy centers in diamond
}

\author{Takuya Isogawa}
\email[]{tisogawa@keio.jp}

\affiliation{Department of Applied Physics and Physico-Informatics, Keio University, Yokohama 223-8522, Japan}

\author{Yuichiro Matsuzaki}
\email[]{matsuzaki.yuichiro@aist.go.jp}
\affiliation{National Institute of Advanced Industrial Science and Technology, Ibaraki 305-8568, Japan}

\author{Junko Ishi-Hayase}
\email[]{hayase@appi.keio.ac.jp}

\affiliation{Department of Applied Physics and Physico-Informatics, Keio University, Yokohama 223-8522, Japan}

%\date{\today}

\begin{abstract}
The measurement of vector magnetic fields with high sensitivity and spatial resolution is important for both fundamental science and engineering applications. In particular, magnetic-field sensing with nitrogen-vacancy (NV) centers in diamond is a promising approach that can outperform existing methods.
Recent studies have demonstrated vector DC magnetic-field sensing with perfectly aligned NV centers, which showed a higher readout contrast than NV centers
having four equally distributed orientations. However, to estimate the azimuthal angle of the target magnetic field with respect to the NV axis in these previous approaches, it is necessary to apply a strong reference DC magnetic field, which can perturb the system to be measured. This is a crucial problem, especially when attempting to measure vector magnetic fields from materials that are sensitive to applied DC magnetic fields.
Here, we propose a method to measure vector DC magnetic fields using perfectly aligned NV centers without reference DC magnetic fields. More specifically, we used the direction of linearly polarized microwave fields to induce Rabi oscillation as a reference and estimated the azimuthal angle of the target fields from the Rabi frequency. We further demonstrate the potential of our method to improve sensitivity by using entangled states to overcome the standard quantum limit. Our method of using a reference microwave field is a novel technique for sensitive vector DC magnetic-field sensing.

\end{abstract}

\pacs{}

\maketitle

\section{\label{Sec.intro}Introduction}
The detection of DC magnetic fields on the nanoscale plays an essential role in a wide range of fields, from biology to condensed matter physics \cite{casola2018probing,schirhagl2014nitrogen}. Substantial effort has been devoted to the development of devices for measuring small magnetic fields in a local region \cite{bending1999local, chang1992scanning, poggio2010force, RevModPhys.89.035002}. The electron spin in the nitrogen-vacancy (NV) center in diamond is 
known to be sensitive to magnetic fields, and it combines the advantages of non-invasiveness, high sensitivity, and nanoscale resolution \cite{maze2008nanoscale, taylor2008high, balasubramanian2008nanoscale, PhysRevLett.107.207210}. The NV center is a spin-1 system, and the frequencies of the ground-state manifolds can be shifted by magnetic fields.
Vector magnetic-field sensing with NV centers has been attracting much attention \cite{maertz2010vector, steinert2010high, pham2011magnetic, tetienne2012magnetic, dmitriev2016concept, sasaki2016broadband}. It has a variety of potential applications, including the imaging of living cells \cite{le2013optical}, mapping of current distribution \cite{nowodzinski2015nitrogen,tetienne2017quantum, ku2020imaging,vool2021imaging}, and reconstruction of vector magnetic fields from ferromagnetic materials \cite{tetienne2015nature, dovzhenko2018magnetostatic}.

The sensing of DC magnetic fields is conventionally performed using 
Ramsey interferometry or optically detected magnetic resonance (ODMR) \cite{ramsey1950molecular, neumann2009excited}. Using different NV orientations, the vector components of the magnetic field along each quantization axis can be used to reconstruct DC or AC vector fields \cite{maertz2010vector, pham2011magnetic}. However, one drawback of these methods is the reduction of the fluorescence contrast. The contrast decreases by a factor of four compared to that of a single NV, which degrades the sensitivity. Multi-frequency control of the NV centers has been proposed and demonstrated as a solution to this problem, but complex microwave pulse sequences are required to implement these schemes~\cite{kitazawa2017vector, yahata2019demonstration, schloss2018simultaneous, zhang2018vector}.
For more practical applications, previous studies devised an ingenious method of measuring vector magnetic fields using a single NV center or perfectly aligned NV centers \cite{chen2013vector, liu2019nanoscale,zheng2020microwave, qiu2021nuclear, wang2021nanoscale, tsukamoto2021vector}.
It is known that the longitudinal and transverse components of the target vector DC magnetic fields are estimated from the shifts in frequencies of the transitions between ground-state manifolds \cite{balasubramanian2008nanoscale}, whereas a reference DC field is used to estimate the azimuthal angle of the target field with respect to the NV axis \cite{liu2019nanoscale, tsukamoto2021vector}. However, the reference DC field can be invasive when measuring the magnetic properties of ferromagnets and superconductors that are sensitive to DC magnetic fields. Therefore, it is desirable to measure the azimuthal angle without a reference DC field for the investigation of various phenomena of magnetic-field-sensitive materials \cite{tetienne2015nature,thiel2016quantitative, dovzhenko2018magnetostatic,lillie2020laser}.

In this study, we propose a scheme to measure the azimuthal angle of the target magnetic field 
without using a reference DC field. In our scheme, the azimuthal angle of the target magnetic field can be estimated from the Rabi frequency using linearly polarized microwave fields as a reference. The transverse component of the DC field changes the direction of the quantization axis of the NV center. On the other hand, when we apply the resonant microwave field,
only the component perpendicular to the quantization axis contributes to the change in the Rabi frequency \cite{wang2015high}. Thus, the change in the direction
of the quantization axis can be estimated from the Rabi frequency, and thus, the direction of the DC magnetic field can be measured.

In addition, we investigate whether the use of entanglement improves the sensitivity of our scheme. By using an entangled state with $L$ spins to measure magnetic fields, the uncertainty of the estimation decreases by $1/L$, which is called the Heisenberg limit, under ideal conditions,
while the uncertainty decreases by $1/\sqrt{L}$ with $L$ separable spins, which is called the standard quantum limit (SQL) \cite{huelga1997improvement,giovannetti2011advances}. However, because the quantum state is fragile against decoherence, it is unclear whether the entangled sensor shows better sensitivity than the classical one. For certain types of noise, the entangled states can outperform SQL~\cite{matsuzaki2011magnetic,chin2012quantum,dur2014improved,kessler2014quantum,tanaka2015proposed}. In our case, the NV centers are affected by magnetic-field noise. Therefore, we investigate the performance of the entanglement sensor in estimating the azimuthal angle with magnetic noise
in our scheme. Furthermore, we numerically show that our sensing scheme with entanglement has better sensitivity than separable sensors, even under the effect of noise.

The remainder of this paper is organized as follows. In Sec.~\ref{Sec.conventional}, we review the conventional vector magnetic-field sensing scheme with standard Ramsey interference using a reference DC magnetic field. In Sec.~\ref{Sec.Our}, we explain our scheme of vector magnetic-field sensing with Rabi oscillation using a reference microwave field. In Sec.~\ref{Sec.entangle}, we investigate the potential improvement in sensitivity when using entangled states. Finally, Sec.~\ref{Sec.conclusion} concludes the paper.

\section{\label{Sec.conventional}Conventional scheme}

\subsection{Measurement of 
longitudinal and transverse components of target magnetic fields}
Here, we review a conventional scheme that measures the
longitudinal and transverse components of target magnetic fields. The Hamiltonian of the NV center is given as
\begin{eqnarray}
\label{H0}
 H_0=
D \hat{S}_z^2 + \gamma_e B_z \hat{S}_z+\gamma_e B_\perp (\cos{\phi_s}\hat{S}_x
+\sin{\phi_s}\hat{S}_y),
\end{eqnarray}
where $D = 2\pi \times 2.87\,\mathrm{GHz}$ is the zero-field splitting, $\gamma_e = 2\pi \times 28\,\mathrm{GHz/T}$ is the gyromagnetic ratio, and $\{\hat{S}_x, \hat{S}_y, \hat{S}_z\}$ are the 
spin-1 operators, and $\{B_z, B_\perp, \phi_s\}$ are the local magnetic-field components in cylindrical coordinates.
Throughout this paper, we assume that $\hbar=1$. Following Refs.~\cite{balasubramanian2008nanoscale} and~\cite{tsukamoto2021vector}, we consider the characteristic equation given by
\begin{eqnarray}
\label{characteristic}
(D-x)^2 x + \frac{1}{2}\gamma_e^2B^2[D(1-\cos{2\theta}) - 2x]=0,
\end{eqnarray}
where $\theta = \arctan(B_\perp/B_z)$ and $B=\sqrt{B_s^2+B^2_{\perp}}$. If we express the frequency of the ground state as $x_0$, the frequencies of the other energy eigenstates can be expressed as $x_+ = x_0 + \omega_+$ and $x_- = x_0 + \omega_-$. By substituting these expressions into Eq.~\ref{characteristic}, we obtain the following equations:
\begin{eqnarray}
x_0 &=&\frac{1}{3}(2D - \omega_{+} - \omega_{-}),\\
B&=&\sqrt{
\frac{1}{3 \gamma_e^2}(\omega^2_{+} + \omega^2_{-} -\omega_{+} \omega_{-} - D^2)},\\
\sin{\theta} &=&\sqrt{
\frac{-x^3_0 + 2D x^2_0 + (\gamma_e^2 B^2 -D^2)x_0}{D\gamma_e^2 B^2}}.
\end{eqnarray}

Thus, from the transition frequencies $\omega_{\pm}$, which can be measured by Ramsey interferometry or ODMR, we can determine the longitudinal and transverse components of the target magnetic fields.

\subsection{Measurement of the azimuthal angle of the target fields}

Next, we review a conventional scheme that measures the azimuthal angle of the target magnetic field using a DC magnetic field as a reference (Fig.~\ref{fig:schematic}(a)). By applying a reference magnetic field with known amplitude and direction, we can determine the azimuthal angle of the target magnetic field from the change in resonant frequency of the NV centers.
The frequency shift can be measured via Ramsey interferometry. 

Our goal here is to determine the azimuthal angle $\phi_s $. When we apply a reference DC magnetic field perpendicular to the NV axis,
an additional term,
\begin{eqnarray}
\label{Hr}
H_r = \gamma_e B_r (\cos{\phi_r}\hat{S}_x+\sin{\phi_r}\hat{S}_y),
\end{eqnarray}
is added to the system Hamiltonian, where $B_r$ and $\phi_r$ are the amplitude and azimuthal angle of the reference field, respectively. This allows us to measure the change in transition frequency between the ground state $|g\rangle \simeq |m_s=0\rangle$ and another energy eigenstate. By using the perturbation theory for small $\gamma_e B_{\mathrm{tot}}/D$,
such transitions frequencies
are approximately given by \cite{barry2020sensitivity}
\begin{eqnarray}
\label{eq:transitionFreq}
\omega_\pm \simeq D\pm \gamma_e B_{z}+\frac{3}{2}\frac{(\gamma_e B'_\perp)^2}{D},
\end{eqnarray}
where $B'_{\perp}=\sqrt{B_\perp^2+B_r^2+2 B_\perp B_r\cos{(\phi_s -\phi_r)}}$ and $B_{\mathrm{tot}}=\sqrt{B_s^2+B'^2_{\perp}}$. 
%We obtain the maximum (minimum) values for both $\omega_\pm$ when $\phi_r=\phi_s$ ($\phi_r=\phi_s+\pi$). 
We consider the subspace spanned by $|g\rangle$ and the first excited state $|e\rangle$ and define $\Delta= D-\gamma_e B_{z}$. In a rotating frame defined by $U=\exp{ (-i \Delta \hat{S}_z t)}$, the effective Hamiltonian of this subspace is given by
\begin{eqnarray}
  H^{\mathrm{(DC)}}_{\mathrm{eff}}= \frac{\delta \omega}{2}(|e\rangle  \langle e|-|g\rangle  \langle g|),
\end{eqnarray}
where $\delta\omega=3(\gamma_e B'_\perp)^2/2D$ is the frequency shift. We create a superposition $|+\rangle=(|e\rangle+|g\rangle)/\sqrt{2}$ by using a $\pi/2$ pulse, and a relative phase can be accumulated during the interaction time $t$. Then, we perform a projective measurement on $|+_y\rangle=(|e\rangle+i|g\rangle)/\sqrt{2}$ with a probability of $P = (1+\sin{(\delta\omega t)})/2$, which can be constructed using a $\pi/2$ pulse and a subsequent optical readout. 
Thus, we can estimate $\phi_s$ by sweeping $\phi_r$ via the Ramsey scheme. 

If we have an approximate value of the azimuthal angle $\phi_s$, we can rewrite it as $\phi_s=\phi_{\rm{a}}+\phi'$, where $\phi_{\rm{a}}$ is the approximate value and $\phi'$ is a small difference from the true value. The frequency shift $\delta\omega$ is then given by $\delta\omega = \delta\omega_{\rm{a}} + \delta\omega'$, where $\delta\omega_{\rm{a}}$ is the approximate value and $\delta\omega'$ is a small difference from the true value. We can adjust the interaction time as
\begin{eqnarray}
\label{eq.tau}
\tau = \frac{n\pi}{\delta\omega_{\rm{a}}},
\end{eqnarray}
and we obtain
$P \simeq (1 +\delta\omega'\tau)/2,$ where we assume $\delta\omega'\tau \ll 1$.
By repeating this experiment many times, we can obtain the probability from the experimental results; thus, the value of $\delta\omega$ can be estimated.

The uncertainty of the estimation is given by
\begin{eqnarray}
\label{uncersec}
\delta\phi=\frac{\sqrt{P(1-P)}}{|\frac{dP}{d\phi}|\sqrt{N}}
\end{eqnarray}
where $N$ is the number of repetitions of the experiment and $\phi=\phi_s -\phi_{r}$ denotes the azimuthal angle to be estimated. We assume that the interaction time $\tau$ is much longer than the state preparation time and measurement readout time. In this case, we have $N \simeq T/\tau$, where $T$ is a given time for the
sensing. We can calculate the uncertainty as 
\begin{eqnarray}
\label{uncer.a}
\delta\phi \simeq \frac{1}{|\frac{d\delta\omega}{d\phi}|\sqrt{T\tau}}.
\end{eqnarray}

Now, we consider the magnetic-field sensing under the effect of decoherence. The dominant noise is due to small fluctuations in the magnetic field due to nitrogen impurities or environmental nuclear spins, which lead to fluctuations in the energy splitting \cite{hayashi2018optimization}. With $D \gg \gamma_e B$, the noise from $B_\perp$ is suppressed by a large $D$ and becomes negligible as compared to $B_z$, as expressed in Eq.~\ref{eq:transitionFreq}. Thus, the effect of decoherence can be described by the standard Lindblad-type master equation:
\begin{eqnarray}
\label{master.dc}
 \frac{d\rho(t) }{dt }=-i[H^{\mathrm{(DC)}}_{\mathrm{eff}},\rho(t) ] - \gamma [\hat{S}_z, [\hat{S}_z, \rho(t) ]],
\end{eqnarray}
where $\gamma$ is the decay rate. In this case, the probability of projecting the state onto $|g\rangle$ can be calculated as $P = \mathrm{Tr}[|+_y\rangle \langle +_y|\rho(t)]$. We numerically calculate the probability $P$ with the initial state 
$\rho(0)=|+\rangle \langle +|$,
and we observe a linear dependence of the probability on a small $\delta\omega'$
for $B_\perp$ between $0.1\, \mathrm{mT}$ and $10 \, \mathrm{mT}$.
The probability $P$ is, thus, expressed as
\begin{eqnarray}
\label{conventional.linear}
P \simeq \frac{1}{2}(1 +c_{\tau,\gamma}\delta\omega'),
\end{eqnarray}
where $c_{\tau,\gamma}$ denotes a coefficient calculated using numerical simulations.
By substituting Eq.~\ref{conventional.linear} into Eq.~\ref{uncersec}, we obtain the uncertainty of the estimation: 

\begin{eqnarray}
\delta\phi \simeq \frac{1}{c_{\tau,\gamma} |\frac{d\delta\omega}{d\phi}| \sqrt{T/\tau}}.
\end{eqnarray}

\section{\label{Sec.Our}Our scheme}

\begin{figure}
\centering
\includegraphics[width=8cm]{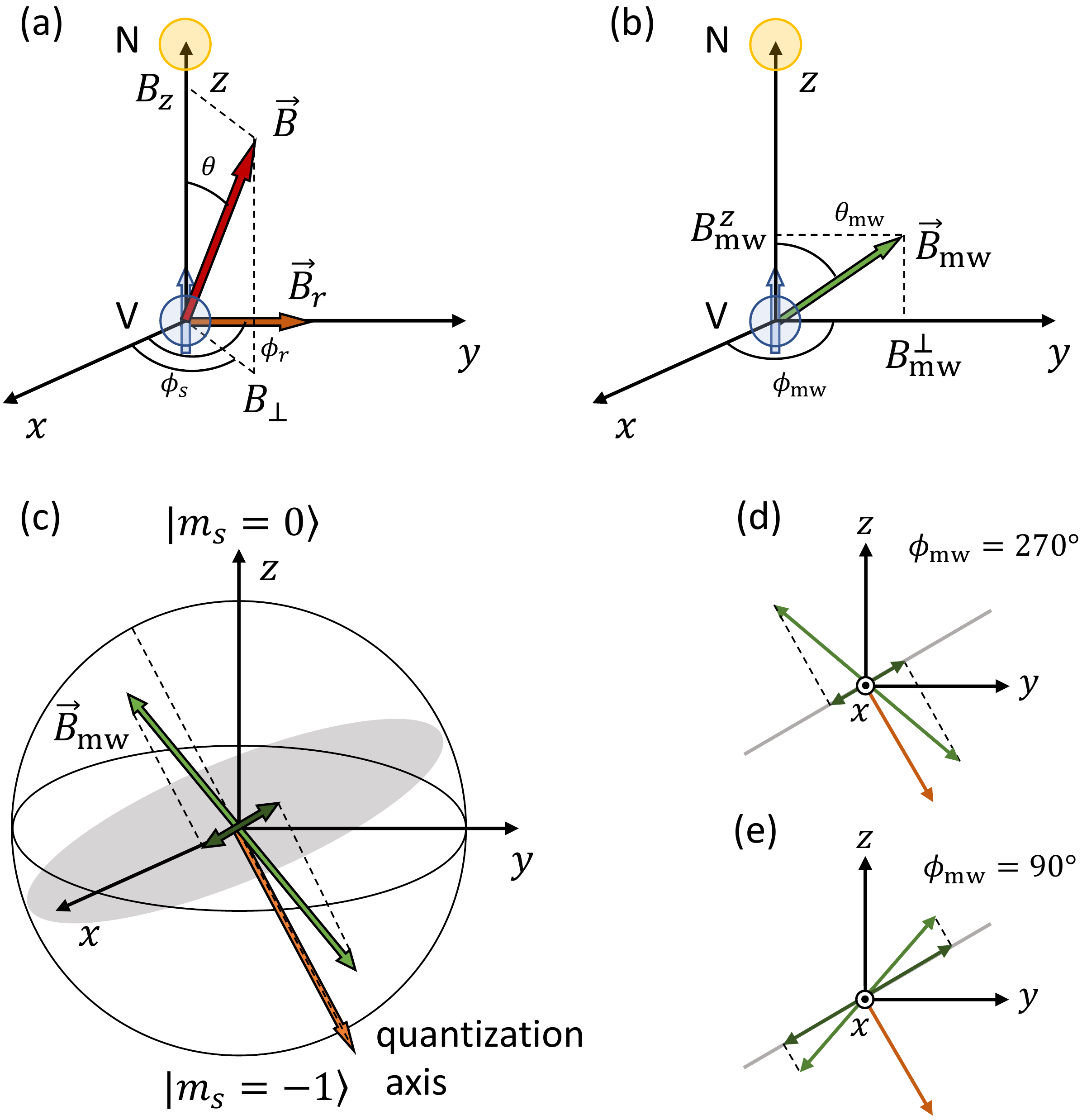}
\caption{Schematic of our vector DC magnetic-field sensing scheme. In (a), we show a configuration of the NV axis, target vector magnetic field, and reference DC magnetic field. 
In (b), we show a configuration of the NV axis and reference microwave field.
In
(c), relations between the quantization axis and direction of the microwave fields in a Bloch sphere are illustrated. In (d) and (e), we show the
Bloch sphere projected onto the y and z planes at $\phi_{\mathrm{mw}}=270^{\circ}$ and $\phi_{\mathrm{mw}}=90^{\circ}$, respectively, with $\phi_s=90^{\circ}$.}
\label{fig:schematic}
\end{figure}

\begin{figure*}
\centering
\includegraphics[width=15.5cm]{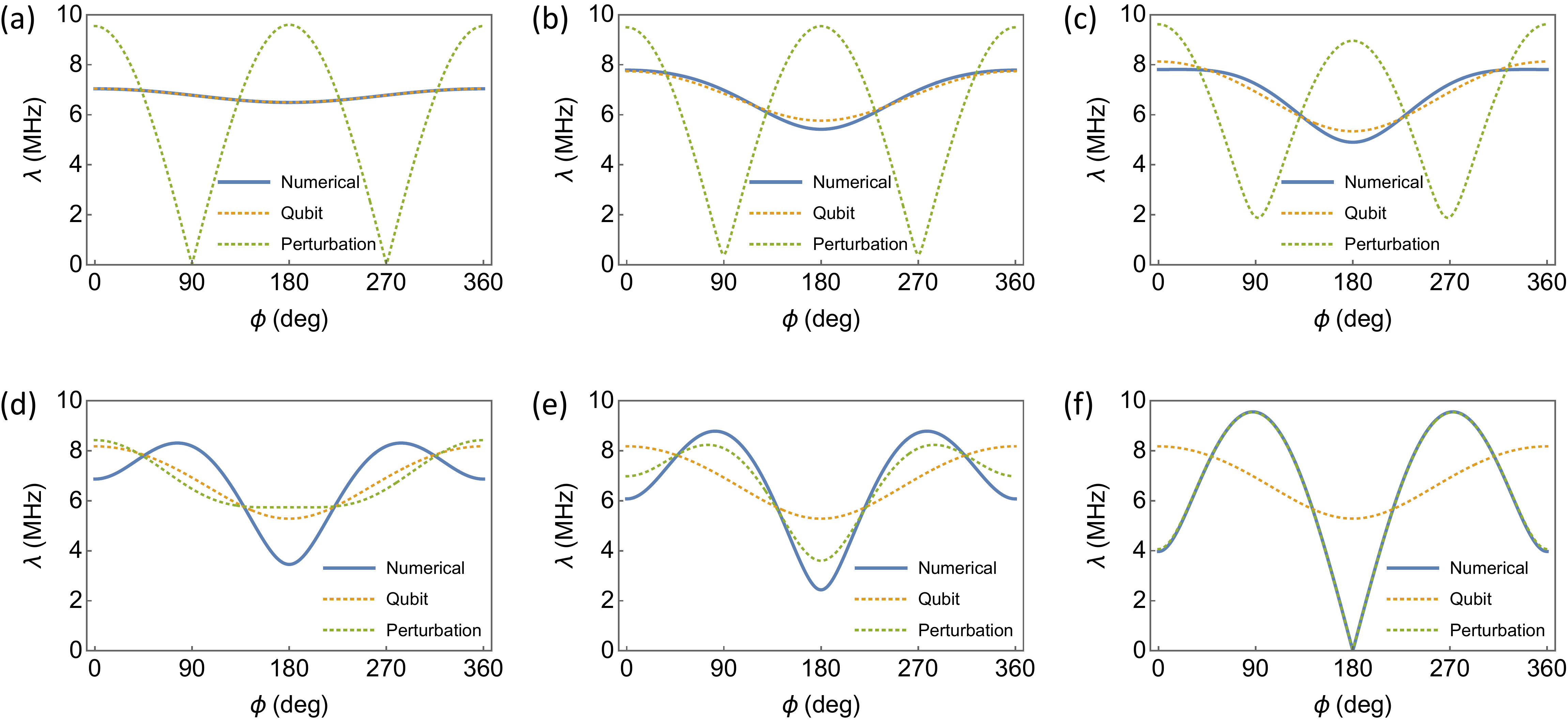}
\caption{
Rabi frequency as a function of the azimuthal angle $\phi$ 
between
the target magnetic field of $B= 8\,\mathrm{mT}$ and the driving microwave field
of $B_{\mathrm{mw}}=1\,\mathrm{mT}$. We set the polar angle of the microwave fields as $\theta_{\mathrm{mw}}=20^\circ$. The polar angles of the target field $\theta$ are (a) $10^\circ$, (b) $40^\circ$, (c) $70^\circ$, (d) $85^\circ$, (e) $87^\circ$, and (f) $89^\circ$. We use the
 numerical (blue), qubit-approximation (orange), and perturbation (green) values, given by Eq.~\ref{exactRabi}, Eq.~\ref{twoLevelRabi}, and Eq.~\ref{perturbationRabi}, respectively.}
\label{fig:Rabi}
\end{figure*}

\begin{figure}
\centering
\includegraphics[width=8.5cm]{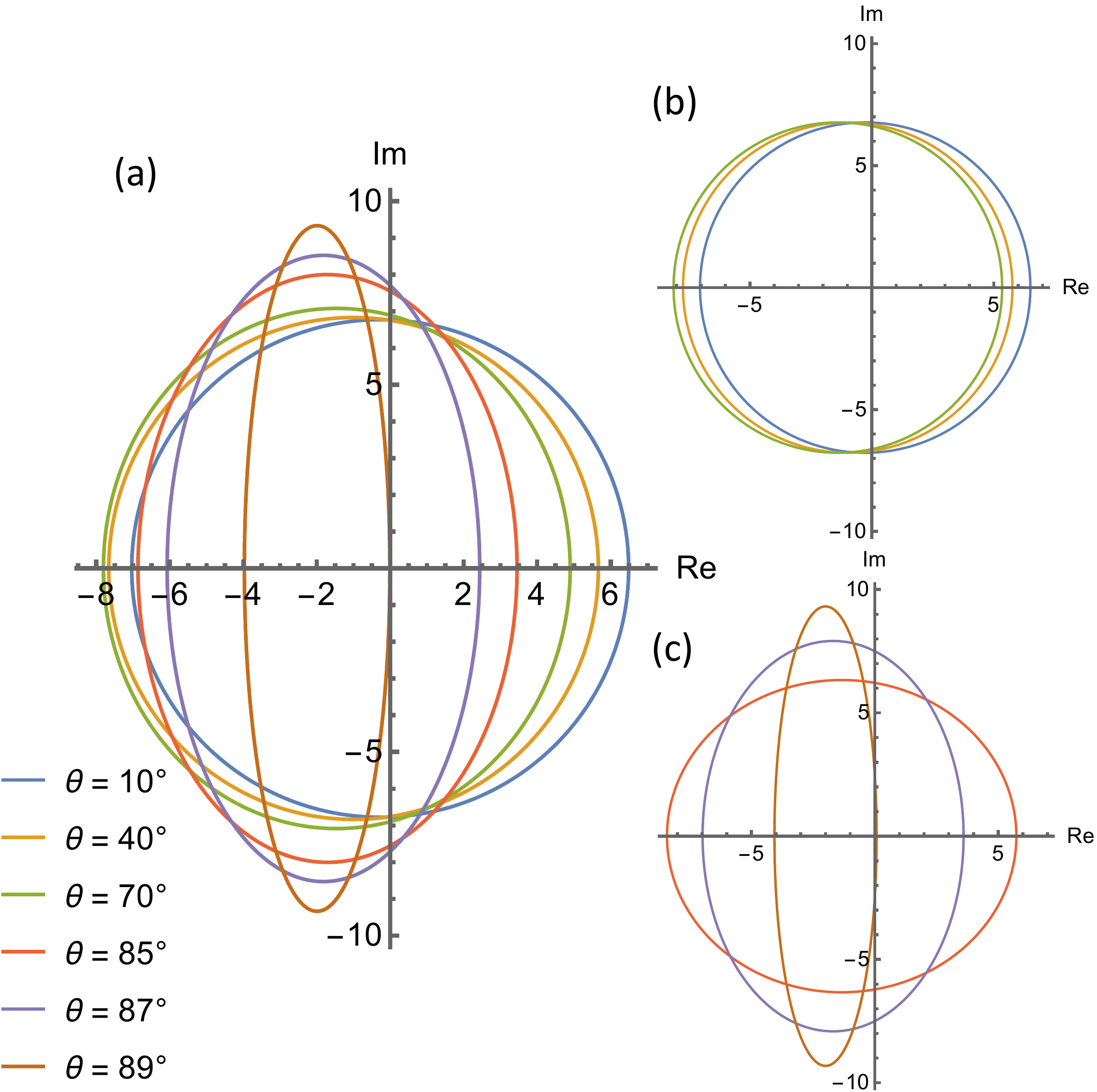}
\caption{Plot of $\langle g|\gamma_e \vec{B}_\mathrm{mw}\cdot\vec{S}|e\rangle$ in Eq.~\ref{exactRabi} on the complex plane. Each ellipse is
obtained by sweeping the azimuthal angle $\phi$ between the target
magnetic field and driving microwave field. The parameters are the same as those used in Fig.~\ref{fig:Rabi}. In (a), we numerically plot $\langle g|\gamma_e \vec{B}_\mathrm{mw}\cdot\vec{S}|e\rangle$ without approximations. In (b), we use the qubit approximation. In (c), we use the perturbation theory described in Eq \ref{perym}.
}
\label{fig:complane}
\end{figure}

Here, we explain our scheme for measuring the azimuthal angles of the target fields without reference DC magnetic fields.
In analogy to the conventional method using the DC magnetic field as a reference, we apply the reference microwave field, the amplitude and direction of which are known (Fig.~\ref{fig:schematic}(b)). The azimuthal angle of the target magnetic field can be estimated by measuring the Rabi oscillations caused by the reference microwave field.

To illustrate our idea,
we start with a simple model by considering a subspace spanned by $|m_s=0\rangle$ and $|m_s=-1\rangle$, which we call the qubit approximation.
The Hamiltonian for this subspace is 
\begin{eqnarray}
 H^{(1/2)}_0=
- \frac{\Delta}{2}\hat{\sigma}_z+\frac{\gamma_e B_\perp}{\sqrt{2}} (\cos{\phi_s}\hat{\sigma}_x
+\sin{\phi_s}\hat{\sigma}_y),
\end{eqnarray}
 where $\{\hat{\sigma}_x,\hat{\sigma}_y,\hat{\sigma}_z\}$ are the 
 Pauli matrices. The quantization axis is along the direction $(\frac{\gamma_e B_\perp}{\sqrt{2}} \cos{\phi_s},\frac{\gamma_e B_\perp}{\sqrt{2}}
\sin{\phi_s},- \frac{\Delta}{2})$.
Any state of a two-level system can be represented by a Bloch sphere (Fig.~\ref{fig:schematic}(c)). 
By diagonalizing $H^{(1/2)}_0$, we have $H^{(1/2)}_0 = \omega^{(1/2)}(|g^{(1/2)}\rangle  \langle g^{(1/2)}|-|e^{(1/2)}\rangle  \langle e^{(1/2)}|)/2$, where 
\begin{eqnarray}
&\omega^{(1/2)}=\sqrt{\Delta^2+2\gamma^2_e B^2_\perp}, & 
\end{eqnarray}
and
\begin{eqnarray}
|g^{(1/2)}\rangle =   c_g \left(\frac{e^{-i\phi_s}(\Delta+\sqrt{\Delta^2+2\gamma^2_e B^2_\perp})}{\sqrt{2}\gamma_e B_\perp} |m_s=0\rangle \right. && \nonumber \\
\left.+ |m_s=-1\rangle\right),  &&  \nonumber\\
|e^{(1/2)}\rangle =  c_e \left( \frac{e^{-i\phi_s}(\Delta-\sqrt{\Delta^2+2\gamma^2_e B^2_\perp})}{\sqrt{2}\gamma_e B_\perp} |m_s=0\rangle \right. && \nonumber  \\
 \left.+ |m_s=-1\rangle\right)  &&\nonumber
\end{eqnarray}
are the eigenstates. Here, $c_g$ and $c_e$ denote normalization constants. When the DC field is applied in a direction different from that of the NV axis, the quantization axis is slightly tilted owing to the perpendicular component of the DC field. Let us consider the Hamiltonian when we apply the microwave field. When microwaves are applied, the Hamiltonian becomes $H^{(1/2)}=H^{(1/2)}_0+H^{(1/2)}_{\mathrm{mw}}$, where $H^{(1/2)}_{\mathrm{mw}}$ is the driving microwave Hamiltonian 
expressed as

 \begin{eqnarray}
 && H^{(1/2)}_{\mathrm{mw}} = \left(\frac{\gamma_e B^z_{\mathrm{mw}}}{2} \hat{\sigma}_z \right.\nonumber \\
 && \left.+\frac{\gamma_e B^{\perp}_{\mathrm{mw}}}{\sqrt{2}}(\cos{\phi_{\mathrm{mw}}}\hat{\sigma}_x+\sin{\phi_{\mathrm{mw}}}\hat{\sigma}_y)\right)\cos{\omega^{(1/2)}t},
 \end{eqnarray}

and we define $\theta_{\mathrm{mw}} = \arctan(B^{\perp}_{\mathrm{mw}}/B^{z}_{\mathrm{mw}})$. In a rotating frame defined by $U=\exp{ (-i H^{(1/2)}_0 t)}$, the effective Hamiltonian is
 \begin{eqnarray}
  H^{(1/2)}_{\mathrm{eff}}=\frac{\lambda}{2}(|g^{(1/2)}\rangle  \langle e^{(1/2)}|+|e^{(1/2)}\rangle  \langle g^{(1/2)}|),
\end{eqnarray}
where
\begin{eqnarray}
\label{twoLevelRabi}
\lambda=
c_g c_e \left|-\gamma_e B^z_\mathrm{mw}+ \frac{\Delta}{B_{\perp}}B^\perp_{\mathrm{mw}}[\cos{(\phi_s-\phi_{\mathrm{mw}})}\right.&&\nonumber\\
\left.+i\sqrt{1+2(\gamma_e B_\perp/\Delta)^2}\sin{(\phi_s-\phi_{\mathrm{mw}})}]\right|&&
\end{eqnarray}
is the Rabi frequency between $|g^{(1/2)}\rangle$ and $|e^{(1/2)}\rangle$. Here, we used rotating wave approximation (RWA). We plot $\lambda$ of Eq.~\ref{twoLevelRabi} against $\phi = \phi_s - \phi_{\mathrm{mw}}$ in Fig.~\ref{fig:Rabi} (orange). Because $\lambda$ is given by the absolute value of a complex number, we can interpret 
this
as the distance from the origin of the complex plane (Fig.~\ref{fig:complane}). On the complex plane, $\lambda$ is the distance between the origin and a point on an ellipse having a width, height, and center of $c_g c_e \Delta B^\perp_{\mathrm{mw}}/B_{\perp}, c_g c_e(\Delta B^\perp_{\mathrm{mw}}/B_{\perp})\sqrt{1+2(\gamma_e B_\perp/\Delta)^2}$, and $(c_g c_e\gamma_e B^z_\mathrm{mw},\,0)$, respectively (Fig.~\ref{fig:complane}(b)). Because $2(\gamma_e B_\perp/\Delta)^2 \ll 1$, the ellipse is close to a circle. If $B^z_\mathrm{mw}$ is sufficiently large, the change in $\lambda$ also becomes large, resulting in better sensitivity. This results from the fact that in the Bloch-sphere description, the Rabi frequency is proportional to the microwave component perpendicular to the quantization axis.

\begin{figure*}
\centering
\includegraphics[width=16cm]{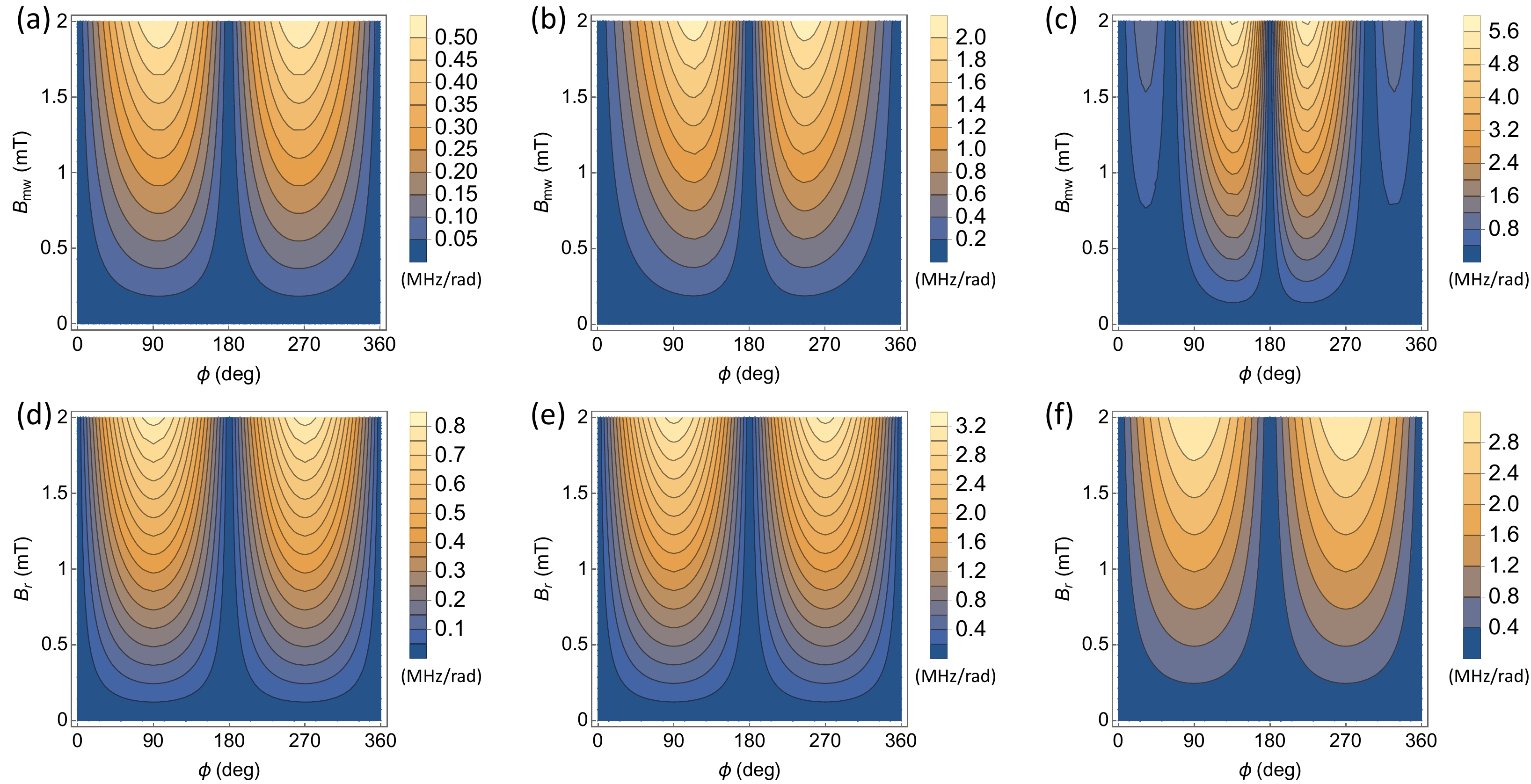}
\caption{(a)--(c) The partial derivative of the Rabi frequency $|\frac{d\lambda}{d\phi}|$ with respect to the amplitude of the microwave field $B_{\mathrm{mw}}$ and azimuthal angle $\phi=\phi_s -\phi_{\mathrm{mw}}$. (d)--(f) The partial derivative of the resonant frequency $|\frac{d\delta\omega}{d\phi}|$ with respect to the amplitude of the reference DC field $B_r$ and azimuthal angle $\phi=\phi_s -\phi_r$. The polar angles of the target field $\theta$ are (a) $10^\circ$, (b) $40^\circ$, (c) $80^\circ$, (d) $10^\circ$, (e) $40^\circ$, and (f) $80^\circ$. Here, $B = 8\,\mathrm{mT}$, $B_{\mathrm{mw}}= 1.0 \,\mathrm{mT}$, $B_{\mathrm{r}}=1.0\,\mathrm{mT}$, and $\theta_\mathrm{mw} = 20^\circ$.}
\label{fig:sensitivity}
\end{figure*}

\begin{figure*}
\centering
\includegraphics[width=15.3cm]{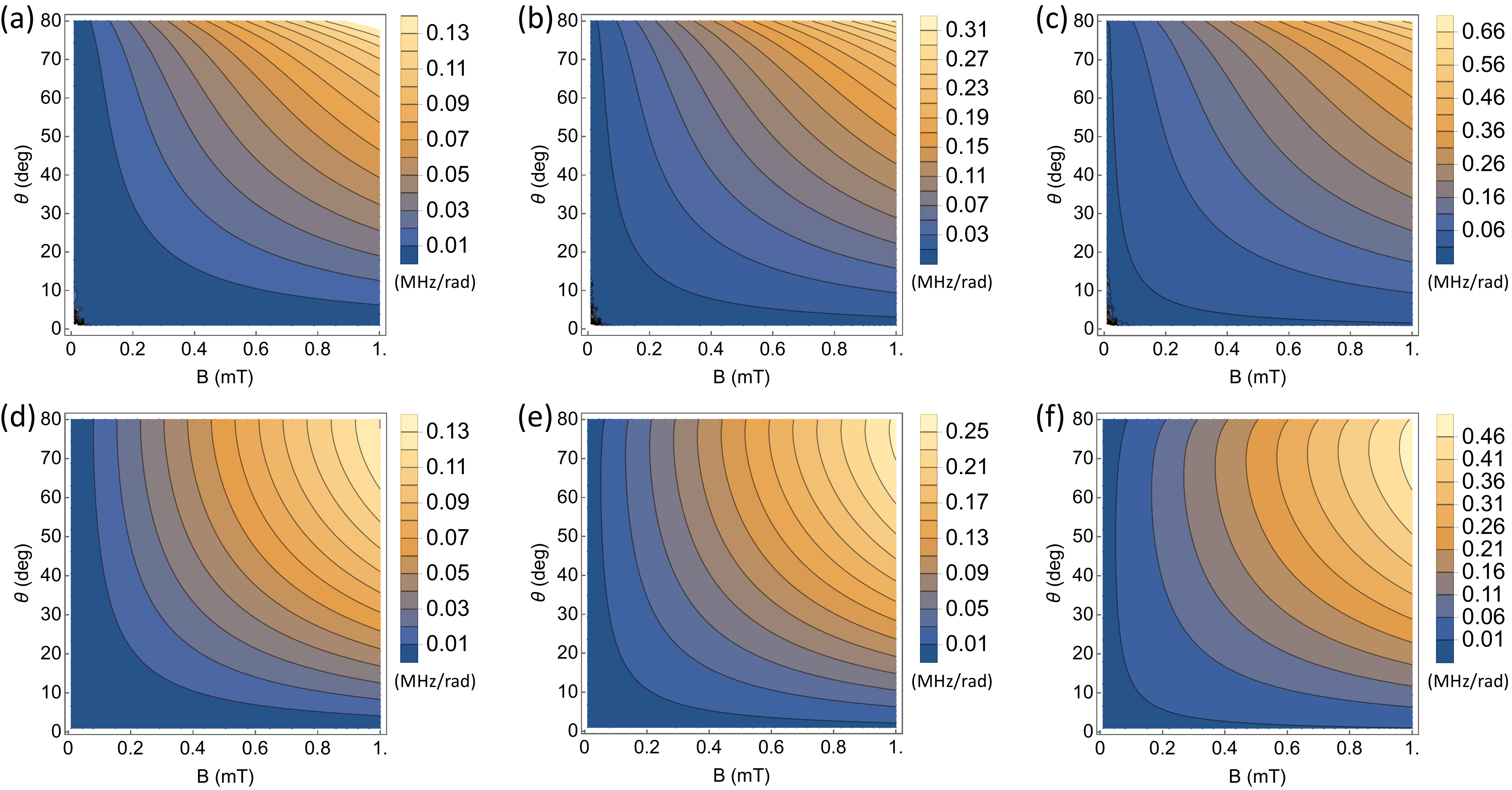}
\caption{(a)--(c) The partial derivative of the Rabi frequency, $|\frac{d\lambda}{d\phi}|$, with respect to the amplitude and polar angle of the target field $B$. (d)--(f) The partial derivative of the resonant frequency, $|\frac{d\delta\omega}{d\phi}|$, with respect to the amplitude and polar angle of the target field $B$. Here, we use the azimuthal angle $\phi$ to maximize the derivative, while (a) $B_{\mathrm{mw}}= 0.5\,\mathrm{mT}$, (b) $B_{\mathrm{mw}}= 1.0\,\mathrm{mT}$, (c) $B_{\mathrm{mw}}= 2.0\,\mathrm{mT}$, (d) $B_{\mathrm{r}}=0.5\,\mathrm{mT}$, (e) $B_{\mathrm{r}}=1.0\,\mathrm{mT}$, (f) $B_{\mathrm{r}}=2.0\,\mathrm{mT}$, and $\theta_\mathrm{mw} = 20^\circ$.}
\label{fig:dynamicrange}
\end{figure*}

\begin{figure}
\centering
\includegraphics[width=7.7cm]{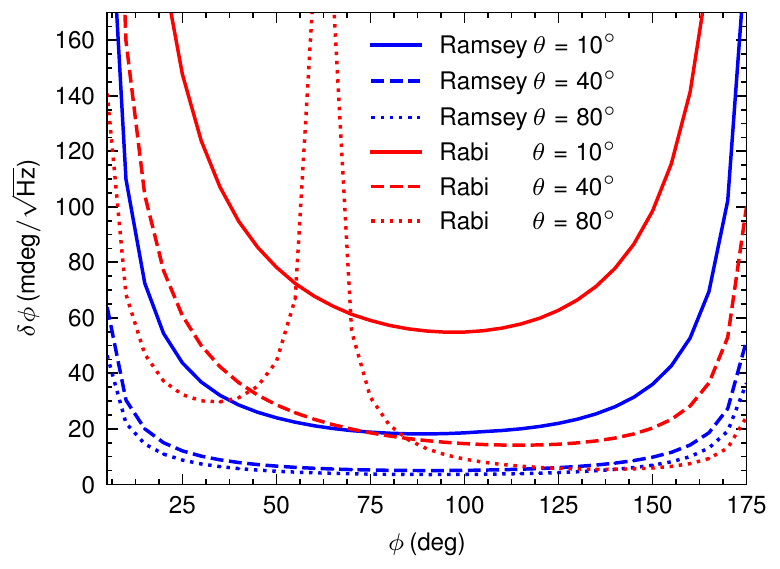}
\caption{Uncertainty of the conventional scheme (blue) and our scheme (red) as a function of the azimuthal angle $\phi=\phi_s -\phi_{\mathrm{mw}}$ for $\theta = 10^\circ$(solid), $\theta = 40^\circ$(dashed), and $\theta = 80^\circ$(dotted). We calculate the uncertainty using Eq.~\ref{uncersec} for $\tau = (2n-1)\pi/2\lambda_{\rm{a}}$ and plot the minimum value. Here, $B = 8\,\mathrm{mT}$, $B_\mathrm{mw} = 1\,\mathrm{mT}$, $B_\mathrm{r} = 1\,\mathrm{mT}$, $\theta_\mathrm{mw} = 20^\circ$, and $\gamma=1.0\,\mathrm{MHz}$.}
\label{fig:SensVsPhi}
\end{figure}

\begin{figure}
\centering
\includegraphics[width=7.7cm]{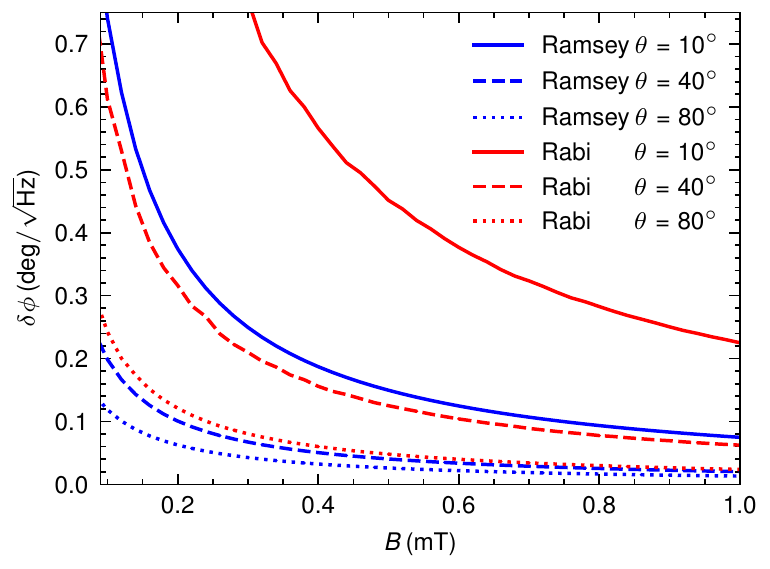}
\caption{Uncertainty of the conventional scheme (blue) and our scheme (red) as a function of the amplitude of the target field $B$ for $\theta = 10^\circ$(solid), $\theta = 40^\circ$(dashed), and $\theta = 80^\circ$(dotted). 
We calculate the uncertainty using Eq.~\ref{uncersec} for $\tau = (2n-1)\pi/2\lambda_{\rm{a}}$ and plot the minimum value. Here, we use the azimuthal angle $\phi=\phi_s -\phi_{\mathrm{mw}}$ to minimize the uncertainty and fix $B_\mathrm{mw} = 1\,\mathrm{mT}$, $B_\mathrm{r} = 1\,\mathrm{mT}$, $\theta_\mathrm{mw} =$ $10^\circ$ (solid), $20^\circ$ (dashed), $20^\circ$ (dotted), and $\gamma=1.0\,\mathrm{MHz}$. }
\label{fig:SensVsB}
\end{figure}

We illustrate this point in Fig.~\ref{fig:schematic}. This
shows the case when the azimuthal angles of the quantization axis are (d) equal to or
(e) equal but opposite in sign to that of the applied-microwave-field direction.
 As can be observed from geometrical interpretation, when the azimuthal angles of the quantization axis and 
 the microwave field
 are equal (equal but opposite in sign), the component of the microwave perpendicular to the quantization axis is minimized (maximized).

Let us consider a more general model: the spin-1 Hamiltonian.
We define $|g\rangle$ and $|e\rangle$ as the ground state and first excited state of the Hamiltonian $H_0$.
When a resonant microwave is applied, the Hamiltonian becomes $H=H_0+H_{\mathrm{mw}}$, where $H_{\mathrm{mw}}$ is the driving microwave Hamiltonian,
\begin{eqnarray}
\label{Hmw}
&&H_{\mathrm{mw}}= \gamma_e \vec{B}_\mathrm{mw}\cdot\vec{S}\cos{\omega_{-}t},\nonumber\\
&&\vec{B}_\mathrm{mw}=(B^z_\mathrm{mw}, B^{\perp}_{\mathrm{mw}} \cos{\phi_{\mathrm{mw}}}, B^{\perp}_{\mathrm{mw}} \sin{\phi_{\mathrm{mw}}}).
\end{eqnarray}
In a rotating frame defined by $U=\exp{ (-i H_0 t)}$, the effective Hamiltonian is
 \begin{eqnarray}
  H^{\mathrm{(MW)}}_{\mathrm{eff}}=\frac{\lambda}{2}(|g\rangle  \langle e|+|e\rangle  \langle g|),
\end{eqnarray}
where 
\begin{eqnarray}
\label{exactRabi}
  \lambda=|\langle g|\gamma_e \vec{B}_\mathrm{mw}\cdot\vec{S}|e\rangle|
\end{eqnarray}
is the Rabi frequency between $|g\rangle$ and $|e\rangle$. Here, we have used the RWA. We cannot obtain a simple analytical form of the eigenstates of $H_0$. Instead, we perform numerical simulations to calculate the Rabi frequency in Fig.~\ref{fig:Rabi} (blue) and Fig.~\ref{fig:complane}(a).
As can be observed in Fig.~\ref{fig:Rabi}, when $\theta$ approaches 90$^{\circ}$, the difference between the results with numerical simulations
and those with qubit approximation becomes more pronounced because the eigenstates become superpositions of all three states.

To understand this point, we use perturbation theory to calculate the Rabi frequency under the assumption that $D$ and $B_\perp$ are much larger than $B_z$. In this case, the eigenstates of the Hamiltonian are expressed as
$|g\rangle \simeq |m_s=0\rangle -\epsilon|B\rangle$, $|d\rangle = |D\rangle$, and $|b\rangle \simeq |B\rangle +\epsilon|m_s=0\rangle$, where $\epsilon = \gamma_e B_\perp/D$, $|B\rangle = (|m_s=+1\rangle +|m_s=-1\rangle)/\sqrt{2}$, and $|D\rangle =( |m_s=+1\rangle -|m_s=-1\rangle)/\sqrt{2}$. 
We call $|B\rangle$ ($|D\rangle$) a bright (dark) state. This representation is particularly useful when we apply a magnetic field orthogonal to the NV axis or electric (strain) fields \cite{doherty2012theory,matsuzaki2016optically,yamaguchi2019bandwidth,yang2020vector}. The first-order corrections to $|g\rangle$ and $|d\rangle$ are 
 \begin{eqnarray}
 |g^{(1)}\rangle=c_{gd}|d\rangle,\quad|d^{(1)}\rangle=c_{dg}|g\rangle+c_{db}|b\rangle ,\nonumber\\
 c_{gd}=-c_{dg}=-\frac{\epsilon\gamma_e B_z}{D+\epsilon\gamma_e B_\perp},\quad c_{db}=-\frac{ B_z}{\epsilon B_\perp}\label{perym},
\end{eqnarray}
which allow us to obtain the Rabi frequency between $|g'\rangle=c_{g'}(|d\rangle+|d^{(1)}\rangle)$ and $|d'\rangle=c_{d'} (|g\rangle+|g^{(1)}\rangle)$,
\begin{eqnarray}
\label{perturbationRabi}
&&\left|\langle d'|H_{\mathrm{mw}}|g'\rangle\right|\nonumber \simeq c_{g'} c_{d'}|- \epsilon\gamma_e B^z_{\mathrm{mw}}\nonumber\\
&&+\gamma_e B^{\perp}_{\mathrm{mw}}[ c_{db}\cos{(\phi_s-\phi_\mathrm{mw})}+i\sin{(\phi_s-\phi_\mathrm{mw})}]|,
\end{eqnarray}
where $c_{g'}$ and $c_{d'}$ are normalization constants. Here as well, we have used the RWA. We also plot $\left|\langle d'|H_{\mathrm{mw}}|g'\rangle\right|$ of Eq.~\ref{perturbationRabi} against $\phi = \phi_s - \phi_{\mathrm{mw}}$ in Fig.~\ref{fig:Rabi} (green). As in the case of the qubit system approximation, we can interpret this as a distance on the complex plane (Fig.~\ref{fig:complane}). In this case, 
the width, height, and center of the ellipse
are $c_{g'} c_{d'} c_{db} \gamma_e B^{\perp}_{\mathrm{mw}}, c_{g'} c_{d'}\gamma_e B^{\perp}_{\mathrm{mw}}$, and $(- c_{g'} c_{d'} \epsilon \gamma_e B^z_{\mathrm{mw}} ,\,0)$, respectively. Because $\epsilon, c_{db} \ll 1$, the ellipse has a long, narrow shape (Fig.~\ref{fig:complane}(c)). Thus, $\lambda$ is maximized when the point on the ellipse is located either at the top or bottom of the ellipse, that is, $\phi_s-\phi_\mathrm{mw}= 90^\circ, 270^\circ$.

The Rabi oscillation can be probed by measuring the population of $|g\rangle$ with the probability $P = (1+\cos{(\lambda t)})/2.$ Thus, we can estimate $\phi_s$ by sweeping $\phi_\mathrm{mw}$ via the measurement of Rabi frequencies.

We can calculate the uncertainty of the estimation in the same manner as in the Ramsey scheme. Assuming that we know an approximate value of the azimuthal angle $\phi_s$, we obtain $P \simeq (1+\lambda'\tau)/2$ by adjusting the interaction time $\tau$, where $\lambda'=\lambda-\lambda_{\rm{a}}$ is a small difference between the approximate value $\lambda_{\rm{a}}$ and the true value $\lambda$. Thus, the uncertainty is
\begin{eqnarray}
\label{uncer.a}
\delta\phi \simeq \frac{1}{|\frac{d\lambda}{d\phi}|\sqrt{T\tau}}
\end{eqnarray}
for a small $\lambda'$, where $\phi=\phi_s -\phi_{\mathrm{mw}}$ denotes the azimuthal angle to be estimated.
Now, we consider the magnetic-field sensing under the effect of decoherence. In this case, the Lindblad-type master equation is given by
\begin{eqnarray}
\label{master.mw}
 \frac{d\rho(t) }{dt }=-i[H^{\mathrm{(MW)}}_{\mathrm{eff}},\rho(t) ] - \gamma [\hat{S}_z, [\hat{S}_z, \rho(t) ]].
\end{eqnarray}
We numerically calculated the probability $P= \mathrm{Tr}[|g\rangle \langle g|\rho(t)]$ with the initial state $\rho(0)=|g\rangle \langle g|$.
We observe a linear dependence of the probability on the small $\lambda'$
for $B_\perp$ between $0.1\, \mathrm{mT}$ and $10 \, \mathrm{mT}$ with $\tau = (2n-1)\pi/2\lambda_{\rm{a}}$. Thus, probability $P$ is expressed as
\begin{eqnarray}
\label{linear.rabi}
P \simeq \frac{1}{2}(1 +\Tilde{c_{\tau,\gamma}}\lambda'),
\end{eqnarray}
where $\Tilde{c_{\tau,\gamma}}$ denotes a coefficient calculated through the numerical simulations.
By substituting Eq.~\ref{linear.rabi} into Eq.~\ref{uncersec}, we obtain the uncertainty of the estimation:
\begin{eqnarray}
\delta\phi \simeq \frac{1}{\Tilde{c_{\tau,\gamma}}|\frac{d\lambda}{d\phi}| \sqrt{T/\tau}}.
\end{eqnarray}

To compare the performance of our scheme with that of the conventional one,
we performed numerical simulations to calculate $|\frac{d\lambda}{d\phi}|$ and $|\frac{d\delta\omega}{d\phi}|$, as shown in Fig.~\ref{fig:sensitivity}. 
When the reference field 
is parallel to the target field, both $\lambda$ and $\delta\omega$ reach
a local maximum or minimum, and the uncertainty $\delta\phi$
tends to infinity at these points.
As can be observed in Fig.~\ref{fig:sensitivity}, when $B_\mathrm{mw}$ is comparable to $B_r$, $|\frac{d\lambda}{d\phi}|$ is of the same order of magnitude as $|\frac{d\delta\omega}{d\phi}|$.
Let us consider a case with a small amplitude of the target field. In this case, $|\frac{d\lambda}{d\phi}|$ and $|\frac{d\delta\omega}{d\phi}|$ decrease 
for a smaller $B_\perp$, as shown in Fig.~\ref{fig:dynamicrange}.

We also obtain $\delta\phi$ by directly calculating Eq.~\ref{uncersec} using the probability $P$ obtained by numerically solving Eqs.~\ref{master.dc} and \ref{master.mw}~\cite{johansson2012qutip}. Figs.~\ref{fig:SensVsPhi} and~\ref{fig:SensVsB} show the dependence of $\delta\phi$ on $\phi$ and $B$, respectively, corresponding to $|\frac{d\lambda}{d\phi}|$
and $|\frac{d\delta\omega}{d\phi}|$ shown in Figs. \ref{fig:sensitivity} and \ref{fig:dynamicrange}. For simplicity, we use the same $\gamma$ for the conventional Ramsey scheme and our Rabi scheme. However, in actual experiments, the coherence time of the Rabi oscillation is usually longer than that of the Ramsey measurements \cite{barry2020sensitivity}. 
Thus, we anticipate that it is experimentally possible to achieve better sensitivity by using our scheme.

\section{\label{Sec.entangle}Sensitivity improvement using entangled states}
Thus far, we described how to measure a vector DC magnetic field with a separable state using a single NV center or aligned NV ensembles along a particular direction. Now, we 
investigate whether a quantum strategy with entanglement
outperforms the classical strategy. It is known that entanglement provides no advantage over separable states in standard Ramsey interferometry in the presence of Markovian dephasing \cite{huelga1997improvement}.
However, Chaves \textit{et al.} showed that entangled states can improve the sensitivity when the transversal noise is dominant, where the direction of the noise is orthogonal to that of the system Hamiltonian \cite{chaves2013noisy,brask2015improved}.
Because our scheme converts the information of the target DC field into the amplitude of the driving microwave, the NV center is mainly affected by transverse noise, unlike the conventional scheme. This allows us to achieve an improvement in the sensitivity by using the entanglement.
For the probe spins, let us consider an ensemble of $L$ NV centers aligned along only one axis. In our study, we assume that the effect of the dipole--dipole interaction between the probe spins is negligible during the Rabi oscillation. This is valid especially when the main noise source is the magnetic noise induced by nitrogen impurities and/or carbon nuclear spins in the environment \cite{bauch2020decoherence,hayashi2020experimental}.

The Hamiltonian of the total system is, thus, given by
\begin{eqnarray}
H&=&\sum_{j=1}^{L}H^{(j)}_0+H^{(j)}_\mathrm{mw},
\nonumber \\
 H^{(j)}_0&=&
(D \hat{S}^{(j)}_z)^2 + \gamma_e B_z \hat{S}^{(j)}_z+\gamma_e B_\perp (\cos{\phi_s}\hat{S}^{(j)}_x
+\sin{\phi_s}\hat{S}^{(j)}_y),\nonumber \\
H^{(j)}_{\mathrm{mw}}&=& \gamma_e \vec{B}_\mathrm{mw}\cdot\vec{S}^{(j)}\cos{\omega_{-}t}.
\end{eqnarray}

We define $|g\rangle_j$ and $|e\rangle_j$ as the ground state and first excited state of the Hamiltonian $H^{(j)}_0$, respectively. In a rotating frame defined by $U=\exp{ (-i \sum_{j=1}^{L}H^{(j)}_0 t)}$, the effective Hamiltonian is 

 \begin{eqnarray}
 \label{Heff_L}
  H_{\mathrm{eff}}= \sum_{j=1}^{L} \frac{\lambda}{2}(|g\rangle  _j\langle e|+|e\rangle _j \langle g|).
\end{eqnarray}

As an initial state, we choose the Greenberger--Horne--Zeilinger (GHZ) state,
\begin{eqnarray}
 |\psi\rangle= \frac{1}{\sqrt{2}}(|+\rangle^{\otimes N}+|-\rangle^{\otimes N}),
\end{eqnarray}
where $|+\rangle=(|g\rangle+|e\rangle)/\sqrt{2}$ and $|-\rangle=(|g\rangle-|e\rangle)/\sqrt{2}$.
As the observable, we choose the parity operator defined by
\begin{eqnarray}
\label{parity}
 \hat{P}_z = \bigotimes^{N}_{j=1} (|g\rangle  _j\langle g|-|e\rangle  _j\langle e|),
\end{eqnarray}
as proposed in Ref.~\cite{brask2015improved}. 

First, let us consider the noiseless case. 
The expectation value of the measurement is given by $P = \cos{(L\lambda t)}$. 
Assuming that we know the approximate value of the azimuthal angle $\phi_s$, we obtain $P \simeq L \lambda' \tau$ by adjusting the interaction time $t$, where $\lambda'$ is a small difference between the approximate and true values. The uncertainty is defined as 
\begin{eqnarray}
\label{uncer.a}
\delta\phi =\frac{\sqrt{1-P^2}}{|\frac{dP}{d\phi}|\sqrt{N}},
\end{eqnarray}
and we obtain $\delta\phi \simeq \frac{1}{L|\frac{d\lambda}{d\phi}|\sqrt{T\tau}}$ for a small $\lambda'$,
where we use $\hat{P}^{2}_{z} = \hat{I}$. 
Thus, we obtain $\delta \phi =\Theta (L^{-1})$, and this scaling is called the Heisenberg limit.

\begin{figure}
\centering
\includegraphics[width=7.7cm]{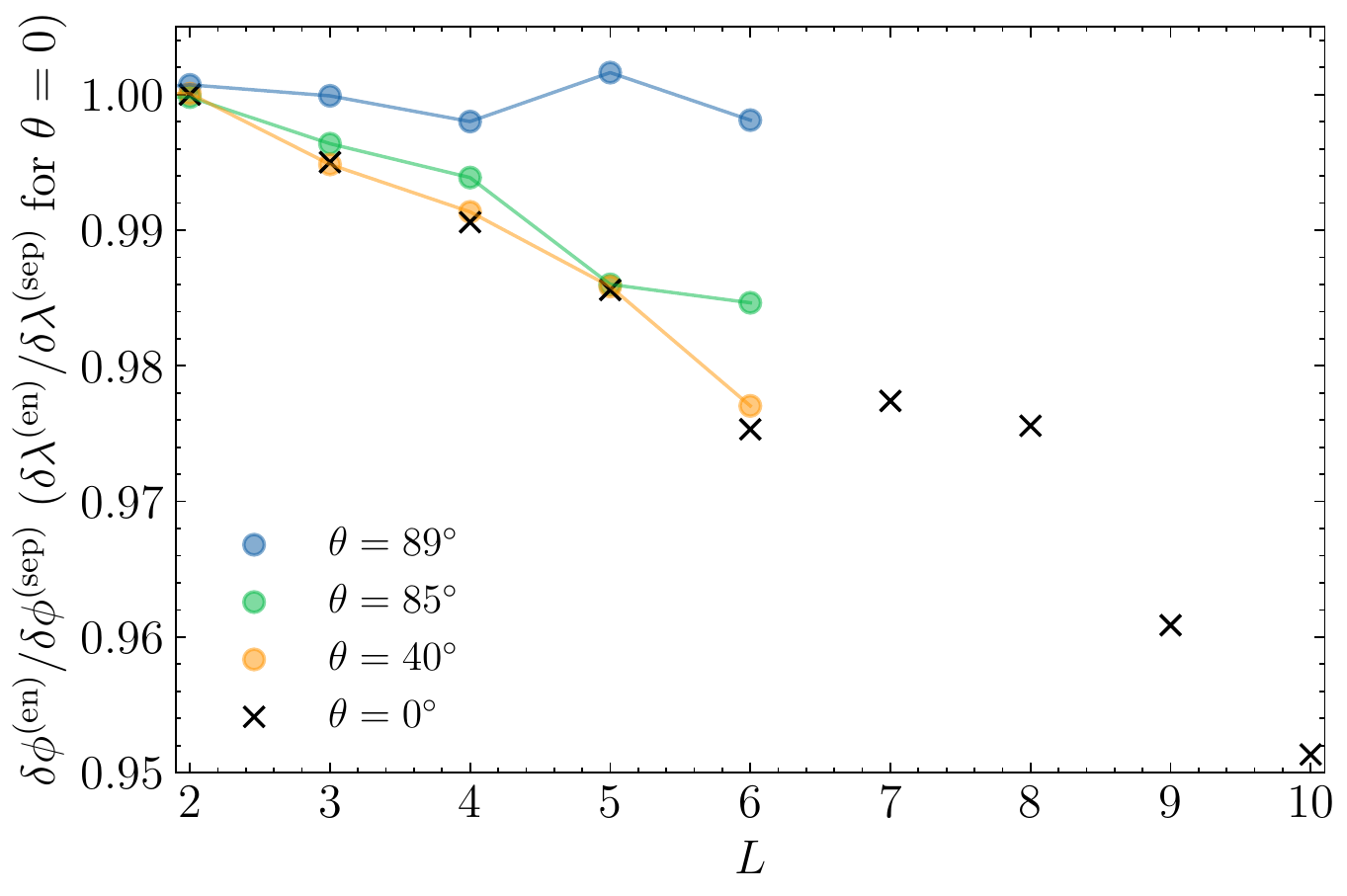}
\caption{Ratio between the uncertainty of the separable sensor and that of the entanglement sensor as a function of $L$. When $\theta$ is not zero, we numerically calculate the uncertainty using Eq.~\ref{uncer.a} for $\tau = (2n-1)\pi/2L\lambda_{\rm{a}}$ and select the minimum value. For $\theta =0$, we plot the uncertainty
based on an analytical solution introduced in \cite{brask2015improved}.
Here, we fix $B = 8 \,\mathrm{mT}$, $B_{\mathrm{mw}} = 1\,\mathrm{mT}$, and $\gamma=1.0\,\mathrm{MHz}$.}
\label{fig:SenSepEn}
\end{figure}

Second, we investigate how the decoherence affects the performance of the entanglement sensor.
To account for the decoherence, we adopt the
Lindblad-type master equation given by
\begin{eqnarray}
\label{master_L}
 \frac{d\rho(t)}{dt}=-i[H_{\mathrm{eff}},\rho(t) ] - \sum_{j=1}^{L}\gamma [\hat{S}^{(j)}_z, [\hat{S}^{(j)}_z, \rho(t) ]].
\end{eqnarray}
 We numerically obtain the expectation value $P= \mathrm{Tr}[ \hat{P}_z \rho(t)]$ with the initial state $\rho(0)= |\psi \rangle \langle \psi |$. We observe a linear dependence of the expectation value on a small $\lambda'$
for $B_\perp$ between $0.1\, \mathrm{mT}$ and $10 \, \mathrm{mT}$ with $\tau = (2n-1)\pi/2L\lambda_{\rm{a}}$. Thus, the expectation value $P$ can be described by
\begin{eqnarray}
\label{linear.ramsey}
P \simeq \Tilde{c_{\tau,\gamma,L}}L\lambda',
\end{eqnarray}
where $\Tilde{c_{\tau,\gamma,L}}$ denotes a coefficient calculated using the numerical simulations.
By substituting Eq.~\ref{linear.ramsey} into Eq.~\ref{uncer.a}, we obtain the uncertainty of the estimation:
\begin{eqnarray}
\delta\phi \simeq \frac{1}{\Tilde{c_{\tau,\gamma,L}}L|\frac{d\lambda}{d\phi}| \sqrt{T/\tau}}.
\end{eqnarray}
Fig.~\ref{fig:SenSepEn} plots the ratio between the uncertainty of the separable sensor and that of the entanglement sensor, up to $L = 6$, when $\theta$ is not zero. We use the same parameters as those used in Fig.~\ref{fig:Rabi}. It should be stated that we cannot define $\delta\phi$ for $\theta =0$. Instead, we calculate $\delta \lambda$ for $\theta =0$ up to $L = 10$ in Fig.~\ref{fig:SenSepEn}, based on an analytical solution introduced in \cite{brask2015improved}, which we will explain below.

It is worth mentioning that our results are a generalization of previous results~\cite{brask2015improved}. For the case of $\theta = 0^\circ$, our model corresponds to that used in~\cite{brask2015improved}, and the uncertainty scales as $\delta \omega =\Theta (L^{-5/6})$ in this case for a large limit of $L$.
On the other hand, when $\theta$ is not zero, the noise operator $\hat{S}_z$ is not completely transverse to the system Hamiltonian, deviating from the model used in ~\cite{brask2015improved}.
In this case, our numerical simulation shows that the improvement due to entanglement decreases as $\theta$ increases.
To understand this point, we use perturbation theory to solve Eq.~\ref{master_L} in the interaction picture. When the effective Hamiltonian given by Eq.~\ref{Heff_L} can be analytically diagonalized, we can calculate the time evolution of $\rho_{\mathrm{I}}(t)$ up to the first order as

\begin{eqnarray}
\label{master_int}
\rho_{\mathrm{I}}(t)\simeq  \rho_{\mathrm{I}}(0)-\gamma \sum_{j=1}^{L} \int_{0}^{t} [\hat{S}^{(j)}_z(t), [\hat{S}^{(j)}_z(t), \rho_{\mathrm{I}}(0)]],
\end{eqnarray}
where $\hat{S}^{(j)}_z(t)
= e^{iH_{\mathrm{eff}}t}\hat{S}^{(j)}_z e^{-iH_{\mathrm{eff}}t}$ with $\gamma t \ll 1$.
Using this expression, we can calculate the expectation value of $\hat{P}_{\mathrm{I}}= e^{iH_{\mathrm{eff}}t}\hat{P_z}e^{-iH_{\mathrm{eff}}t}$.
%for $L=2,\dots,6$.

The analytical formula for $\theta = 0^\circ$ is given up to the third order by

\begin{eqnarray}
\label{expansion}
\mathrm{Tr}[\hat{P}_{\mathrm{I}}\rho_{\mathrm{I}}(t)]
 &=& 1 -\frac{1}{2}L^2\lambda^2 t^2 \nonumber\\
&&+ \frac{1}{6}L(3L-2)\gamma \lambda^2 t^3 + O(t^4),
\end{eqnarray}
which shows a cubic decay against time. 
It is shown that the entanglement strategy outperforms the classical strategy in scaling for such a cubic decay~\cite{brask2015improved}.
On the other hand, for $\theta = 90^\circ$, we obtain
\begin{eqnarray}
\label{expansion}
\mathrm{Tr}[\hat{P}_{\mathrm{I}}\rho_{\mathrm{I}}(t)]= 1-\frac{1}{2}L\gamma t + O(t^2),
\end{eqnarray}
which shows a linear decay. It is known that an entanglement sensor does not have any advantage over separable sensors when the system exhibits a linear decay with time
\cite{huelga1997improvement}. For $0^\circ <\theta < 90^\circ$, we cannot obtain an analytical form of $\mathrm{Tr}[\hat{P}_{\mathrm{I}}\rho_{\mathrm{I}}(t)]$, because we cannot analytically diagonalize the effective Hamiltonian. 
However, we can guess that the sensitivity of the entanglement sensor should be between the two, which is consistent with our numerical results.

\section{\label{Sec.conclusion}Conclusion}
In conclusion, we have proposed a scheme for vector DC magnetic-field sensing that uses microwave fields as a reference to determine the azimuthal angle of the target magnetic field. We have shown that the direction of the DC magnetic field can be measured from the Rabi frequency because the component of the resonant microwave field perpendicular to the quantization axis contributes to driving the Rabi oscillation. In addition, we investigated the potential of our method to improve the sensitivity by using entangled states and found that our sensing scheme with entanglement has better sensitivity than separable sensors, even under the effect of noise for most of the parameters. Our results pave the way for investigating various phenomena of materials that are sensitive to static magnetic fields.

\section{\label{Sec.acknowledgements}Acknowledgements}
The authors gratefully acknowledge discussions with Kento Sasaki and Kensuke Kobayashi. T.~I.~thanks Yasutomo Ota, Guoqing Wang, Pai Peng, Changhao Li, and Paola Cappellaro for their helpful discussions.
This work was supported by MEXT Q-LEAP (No.~JPMXS0118067395), MEXT KAKENHI (No.~18H01502), and CSRN, Keio University. This work was also supported by the Leading Initiative for Excellent Young Researchers MEXT Japan, JST presto (Grant No.~JPMJPR1919) Japan, KAKENHI (20H05661), and NICT Quantum Camp 2021.

\bibliography{aqm}

\end{document}